# SOLAR PHYSICS IN THE 2020s: DKIST, PARKER SOLAR PROBE, AND SOLAR ORBITER AS A MULTI-MESSENGER CONSTELLATION[1]

Valentin Martinez Pillet, NSO, USA;
Alexandra Tritschler, NSO, USA;
Louise Harra, PMOD/WRC & ETH, Switzerland;
Vincenzo Andretta, INAF/OACN, Italy;
Angelos Vourlidas, JHU/APL, USA;
Nour-Eddine Raouafi, JHU/APL, USA;
Ben L. Alterman, SwRI, USA;
Luis Bellot Rubio, IAA, Spain;
Gianna Cauzzi, NSO/INAF, Italy;
Steven R. Cranmer LASP/CU, USA;
Sarah Gibson, HAO/NCAR, USA;
Shadia Habbal, IfA/UH, USA;
Yuan-Kuen Ko, NRL, USA;
Susan T. Lepri, UMich, USA;
Jon Linker, PSI, USA;
David M. Malaspina, LASP/CU, USA;
Sarah Matthews, MSSL/UCL, UK;
Susanna Parenti, IAS, France;
Gordon Petrie, NSO, USA;
Daniele Spadaro, INAF/OACT, Italy;
Ignacio Ugarte-Urra, NRL, USA;
Harry Warren, NRL, USA;
Reka Winslow, UNH, USA.

---

[1] Based on the discussions at the DKIST Critical Science Plan Workshop 4: "Joint Science with Solar Orbiter and Parker Solar Probe". JHU/APL, 13 – 15 March 2018, Laurel, MD, USA

**Abstract**

The National Science Foundation (NSF) Daniel K. Inouye Solar Telescope (DKIST) is about to start operations at the summit of Haleakalā (Hawai'i). DKIST will join the early science phases of the NASA and ESA Parker Solar Probe and Solar Orbiter encounter missions. By combining in-situ measurements of the near-sun plasma environment and detail remote observations of multiple layers of the Sun, the three observatories form an unprecedented multi-messenger constellation to study the magnetic connectivity inside the solar system. This white paper outlines the synergistic science that this multi-messenger suite enables.

1. **The multi-messenger constellation**

The next decade will see the start of the NSF's DKIST (Tritschler et al. 2016) operations coinciding with the first encounter missions aimed at mapping the physical conditions in the vicinity of the Sun, the NASA mission Parker Solar Probe (PSP, launched in August 2018) and the ESA/NASA mission Solar Orbiter (launch in February 2020). DKIST, PSP, and Solar Orbiter (Figure 1) will form an unprecedented solar corona and inner heliospheric campaign targeted at understanding how stars create and control their magnetic environments. DKIST will contribute in many ways to this campaign. A fundamental aspect of the synergistic science is the ability of NSF's DKIST to measure the off-limb solar corona and map its magnetic field quantitatively.

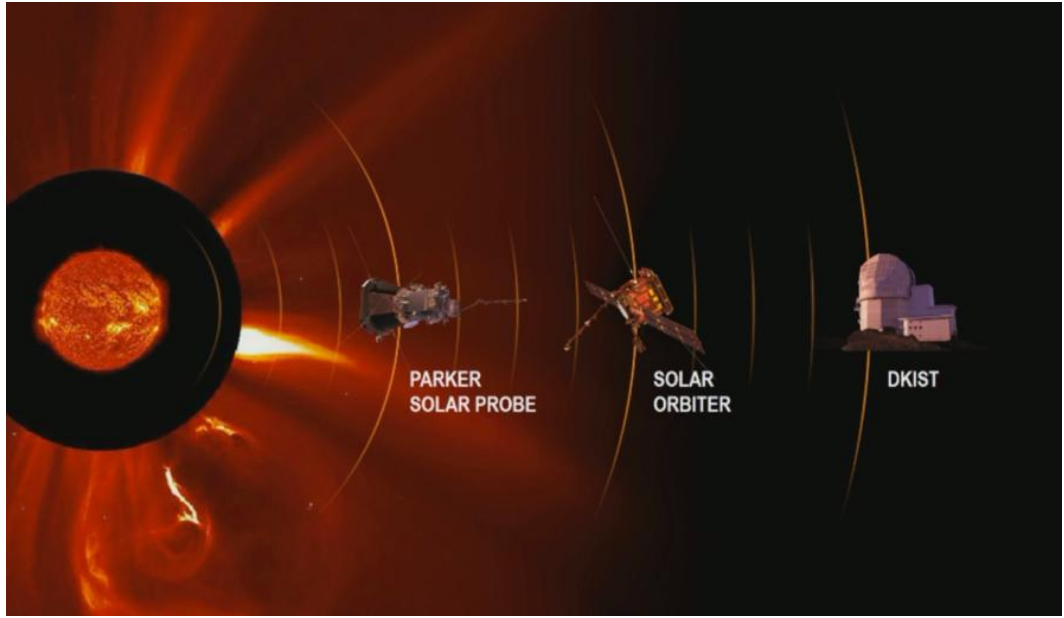

*Figure 1. DKIST, PSP, and Solar Orbiter form a multi-messenger campaign that will revolutionize coronal and heliospheric physics over the next decade. Image courtesy of NASA.*

PSP and Solar Orbiter in-situ instruments measure the plasma particles' kinetic properties, ion composition and elemental abundances, and local electric and magnetic fields. They also have the ability to remote-sense solar radio emission, including source location determination via direction-finding analysis. DKIST and Solar Orbiter perform imaging, spectroscopy, and polarimetry of the solar atmosphere, covering the photosphere, chromosphere, and the corona.  Both PSP and Solar Orbiter image the tenuous corona and heliosphere they fly through using imagers mounted on the side of the spacecraft.  *By combining their data and investigating the cause-effect relationship between the processes observed in the solar atmosphere and their consequences in the solar system, the three observatories will effectively create a multi-messenger decade for solar and heliospheric physics.*

This paper highlights the science enabled by the DKIST's new capabilities in conjunction with PSP and Solar Orbiter.  The National Solar Observatory (NSO), and other ground-based observatories, also contribute to this multi-messenger era with data from synoptic observations such as those performed by GONG (Hill 2018) or the Mauna Loa Solar Observatory coronagraphs, K-Cor and CoMP (Tomczyk et al. 2008). (See Schwadron et al.[2] for a complete discussion of ground-based support for PSP.) Of particular relevance to both space missions are synoptic magnetograms produced by GONG and SDO/HMI that predict the instantaneous magnetic connectivity of the spacecraft during the encounter periods (modeling of the PSP connectivity during its first perihelion is discussed in van der Holst et al. 2019 and Riley et al. 2019; for the Solar Orbiter mission see Rouillard et al. 2020).

## 2. Tracing the Magnetic Connectivity from the Sun into the Heliosphere

Understanding the magnetic connectivity between the planets and the Sun, and by extension between exoplanets and their parent stars, requires tracing particles and fields measured in the heliosphere back to their source regions on the Sun. Establishing the underlying mechanisms responsible for this connectivity has been a daunting task from 1 AU. The exceptions are (i) major Coronal Mass Ejections (CMEs) whose progenitors at the surface are often well-identified thanks to the association of CMEs with flares, dimming effects, and the in-situ detection of the resulting ejected cloud (Gopalswamy et al. 2018), and (ii) relativistic electron beams, which can, for some events, be traced back to their acceleration location in the corona via hard X-ray and radio emission (e.g., Cairns et al., 2018; Krucker et al. 2008). Other, steadier and subtler processes dictating this connectivity, such as the various forms of the solar wind, have not been conclusively linked with processes and evolving configurations on the Sun. The fast solar wind is known to originate in open field regions or coronal holes, and long-lived low-latitude coronal holes are known to drive recurring high-speed solar wind streams at the Earth (e.g., Gibson et al. 2009). However, the specific small-scale structures (flux tubes, etc.) from where it emanates and the underlying physics remain elusive (Cranmer and Winebarger 2019). The mapping of the slow solar wind into specific solar structures is also poorly understood. Its creation is attributed to a variety of configurations, such as the heliospheric current sheet and pseudostreamers (Abbo et al., 2016). Beyond this, our knowledge about the exact location of the processes that give rise to the expanding outer atmosphere of the Sun (see Zhao et al. 2017 and references therein) is rather inconclusive.

At 1 AU, plasma from the Sun has evolved and mixed with other constituents via transport and interactions resulting in conditions that prevent a satisfactory mapping to its solar sources.

---

[2] *https://sppgway.jhuapl.edu/sites/default/files/Pubs/SPP-GBN-WhitePaper-v5.0.pdf*

Similarly, the absence of measurements of the solar wind close to the Sun has challenged the evaluation of competing models of solar wind acceleration. Near-sun solar wind samples should be easier to map with sufficient precision to a solar source region and a specific process. For this reason, over several decades, the solar community has advocated for missions that approach our star and measure the fields and particles closer while they still preserve many of the properties (composition, charge, kinetic properties, etc.) that link them to their solar sources. This proximity to the Sun was the rationale behind the Parker Solar Probe (Fox et al. 2016) mission in the NASA Heliophysics portfolio and the Solar Orbiter mission included in the European Space Agency (ESA; Müller 2013) science program. NASA also participates in the Solar Orbiter mission.

PSP and Solar Orbiter are encounter-class missions instead of observatory-class missions. No observing proposals are submitted to the mission teams. Alternatively, predefined measurements are made during selected windows coinciding with the most scientifically relevant portions of the orbits, in particular the periods of closest proximity to the Sun and largest inclination out of the ecliptic plane.

Both missions use Venus Gravity Assist Maneuvers (GAM) to get closer to the Sun (PSP) or to increase the angle to the plane of the ecliptic (Solar Orbiter), allowing a better view of the solar poles and a better mapping of the 3D heliospheric volume. A mission like Solar Orbiter that escapes the ecliptic to provide imagery of the solar poles can make a profound impact on our understanding of the solar wind formation and acceleration.

The proximity to the Sun achieved by PSP generates corotation opportunities (or quasi-corotation for Solar Orbiter) periods where the spacecraft flies right above the same solar feature for extended periods (hours and days), allowing the study of its evolution and how the properties of the solar wind vary depending on changes in the source regions. These corotation periods help disentangle spatial from temporal variations in the heliospheric structures and establish cause-and-effect relationships between solar processes and in-situ measurements.

As the encounter windows depend on the particular orbit of PSP or Solar Orbiter, they result in varying Earth-spacecraft configurations. For DKIST, the best times for synergistic science occur when Earth is in quadrature with one of the spacecrafts. In that case, off-limb coronal DKIST observations map the inner solar vicinity of the spacecraft's trajectory. However, encounter windows in alignment between the Earth and the satellites allow for additional scientific opportunities, with DKIST observing on the disk or in the off-limb corona.

## 3. PSP and DKIST: Enabling Science by Combining In-Situ Near-Sun Measurements and 1 AU Coronal Observations

PSP was launched in August 2018 and has already performed four perihelia at a distance of 35 $R_\odot$ (0.16 AU), following the first Venus GAM. The nominal mission uses a total of 24 perihelion encounters, with the last one occurring in June 2025. This period includes four years after the expected start of DKIST operations in October 2020. In this period, six additional Venus flybys will bring PSP ever closer to the Sun, reaching 9.86 $R_\odot$ (0.0459 AU) during the last three encounters in 2024/25. PSP will spend about a thousand hours below 20 $R_\odot$, and fifteen hours below 10 $R_\odot$. This inner region is crucial for the scientific objectives of PSP as at this distance, the solar wind is predicted to be sub-Alfvénic, and plasma interactions generate waves that can carry energy back towards the Sun. The coexistence of inward and outward propagating Alfven waves drives turbulence, which in turn is often invoked as a mechanism to heat the solar corona (Cranmer & van Ballegooijen 2005). When the solar wind accelerates and becomes super-Alfvénic, the inwardly directed waves no longer reach back to the corona, and this decoupling is often taken as the definition for the initiation of the heliospheric solar wind. PSP enters this critical region where the solar wind emanates and measures the local plasma conditions.

PSP carries three in-situ instrument suites and a heliospheric imager. The Solar Wind Electrons Alphas and Protons (SWEAP) investigation (Kasper et al. 2016) measures the particles that compose the bulk of the solar wind: electrons, protons, and alpha particles. The four SWEAP sensors measure the velocity distribution functions of the particles with good energy and angular resolution. SWEAP is a critical instrument for connectivity studies as it detects particles that one can trace back to chemical composition on the solar atmosphere. Additional particle measurements at higher energies are performed by the Integrated Science Investigation of the Sun (IS☉IS, McComas, et al. 2016). IS☉IS is the primary instrument to study the particles created by energetic events, such as flares and CMEs.

The FIELDS instrument (Bale et al. 2016) measures DC-coupled magnetic and electric fields, plasma wave magnetic and electric fields, and solar radio emissions. From the measured local electric and magnetic fields, it produces an estimate of the Poynting flux, i.e., the electromagnetic energy flux crossing that region of space. The 3-components of the local magnetic field obtained by FIELDS allow mapping back to the outer corona and solar surface.

Finally, the Wide-Field Imager (WISPR, Vourlidas, et al., 2016) is the only imaging instrument aboard the spacecraft. It measures the white-light scattered by dust (F-Corona) and by electron (Thomson scattering). WISPR looks at the large-scale structure of the corona and solar wind before the spacecraft passes through it. It will image the fine-scale structuring of the solar corona and determine whether a predicted dust-free zone exists near the Sun. At closest approach (in 2024), the inner field-of-view of WISPR sees about two solar radii above the surface, which is close to the area that is observable by the DKIST and its instrumentation, i.e. reaching to 1.5 $R_{\odot}$. This creates a unique opportunity to test whether the diffuse coronal brightness seen in He I 1083 nm can be due to the neutralization of alpha particles by dust as proposed by Moise et al. (2010). This is a prime example of synergistic science, where progress has been difficult thus far, but that can be enabled by coordinated observations from the DKIST's Cryogenic Near IR Spectro-Polarimeter (Cryo-NIRSP), WISPR, and SWEAP.

Clarifying detection of neutral Helium in the million-degree solar corona is a key goal for DKIST because the He I 1083 nm line is the only permitted IR line available, and it has the potential to unlock new coronal magnetic field diagnostics through the Hanle effect as demonstrated by Dima et al. (2016). These authors describe an algorithm that combines linear polarization measurements of the Si X 1430 nm forbidden line, in the saturated Hanle regime for coronal conditions, with polarization observations of the He I 1083 nm permitted line to infer the three components of the coronal magnetic field. Such measurements will constrain the magnetic topology significantly and can feedback into global coronal models, improving the accuracy of the heliospheric magnetic configurations predicted by them. While at solar minimum, coronal-heliospheric models using photospheric synoptic data can produce satisfactory mappings of the connectivity back to the Sun, the more complex corona that PSP will encounter during solar maximum will need the most refined available input data (Raouafi et al., 2016). With this in mind, the DKIST Critical Science Plan is considering a synoptic program that regularly measures the solar corona in anticipation of PSP encounters. The lines and techniques used by this program can change as we learn how to generate the best input coronal magnetic data for the models. Most likely, such a synoptic program will initially use the forbidden Fe XIII line at 1075 nm in a way similar to the CoMP instrument, but will also include the line-of-sight component from the Stokes V signals that are measured by DKIST.

The chemical composition of the solar wind is an indicator of the source region on the Sun (Geiss et al. 1995) and is a critical ingredient for establishing the magnetic connectivity. An example relevant to PSP is the relation between in-situ measurements of Helium at 1 AU and their dependence on the solar wind speed and phase of the activity cycle (Kasper et al. 2007, 2012). Observations show that the smallest Helium abundances are observed during sunspot minimum, with slower solar winds showing a more pronounced correlation than faster solar winds. Indeed, the Helium abundance of the fast solar wind minimally changes as a function of the cycle. These correlations point to mechanisms that affect the second most abundant constituent of the solar plasma with an effectiveness that varies according to the magnetic activity levels. The underlying mechanism remains unknown, but there are indications that high coronal Helium abundance prevents the escape of coronal plasma, which may also explain the Helium-enriched regions found in CME ejecta. The combination of PSP and DKIST over the next decade will provide new ways to investigate this fundamental problem (Alterman and Kasper 2019). Using the SWEAP instrument, PSP will measure the in-situ abundance of Helium much closer to the Sun, providing a less ambiguous identification of the source regions. As the Helium depletion is known to occur lower in the corona (Laming 2015), chromospheric observations will play a crucial role. All DKIST instruments can contribute to this investigation. DKIST can observe on the disk and the off-limb corona several Helium lines, most importantly He I 1083 nm with the Diffraction-Limited Near-Infrared Spectro-polarimeter (DL-NIRSP) (near limb corona) and Cryo-NIRSP up to 1.5 $R_\odot$. In the visible, the Visible Spectro-Polarimeter (ViSP) can access a number of He I and He II spectral lines and observe them in the near-limb region, benefiting from the coronagraphic capabilities of DKIST and the dark sky conditions on Haleakalā. Most likely the relevance of these measurements will become more critical at the closest PSP distances near solar maximum in 2024, but the solar cycle fluctuations described above indicate that synoptic measurements should start as soon as DKIST becomes operational in 2020, in the midst of what is expected to be a deep solar minimum. We plan to define the exact combination of DKIST instruments and lines during the first year of operations (the Operations Commissioning Phase, OCP).

## 4. Solar Orbiter, PSP, and DKIST: A Multi-Messenger Approach to Solve the Magnetic Connectivity Conundrum

Solar Orbiter launched in February 2020 on a NASA Atlas V 411 rocket from Space Launch Complex 41 at Cape Canaveral. The joint ESA/NASA mission combines four in-situ instruments and six remote sensing telescopes to connect observed processes on the Sun with local measurements. Solar Orbiter seeks to understand the magnetic connectivity between the Sun and the heliosphere. As such, it was designed and built as a multi-messenger mission from its conception. With openings in the heat shield to allow the onboard telescopes to observe the Sun, it cannot approach the Sun as close as PSP does. Solar Orbiter reaches a distance to the Sun of 60 $R_\odot$ (0.28 AU), and an inclination of 33° relative to the ecliptic plane at the end of its (extended) mission. It uses a series of 8+8 (nominal plus extended mission) encounter periods, each comprised of three 10-day windows during which the remote sensing instruments operate. In-situ instruments are always operational. The remote sensing windows are centered on the most scientifically relevant parts of the orbit (see Figure 2), perihelion, and maximum and minimum heliographic latitude, both providing vantage points for observing the solar poles. The exact timing of these windows is based on long term planning and may evolve during the mission. The choice of remote sensing windows will also consider coordinating opportunities with PSP. The first opportunity for collaboration among the three facilities will occur after the 6th PSP perihelion passage in September of 2020, a time when DKIST will be ramping-up operations during its OCP.

The Solar Orbiter payload is more complex than the PSP payload, and we defer its description to the ESA/NASA mission's red book[3]. All remote sensing instruments offer multiple opportunities to coordinate with DKIST. The Solar Orbiter Polarimetric and Helioseismic Imager (PHI; Solanki et al. 2020), together with DKIST/ViSP observations of the same spectral line, will provide an opportunity to carry out magnetic field stereoscopy for the first time and unambiguously solve the 180° ambiguity intrinsic to all Zeeman measurements. This stereoscopic study will certainly include other instruments on the Earth-Sun line-of-sight, but the sensitivity to transverse fields that DKIST will provide might prove crucial for satisfactory disambiguation, in particular in the quiet Sun. The Extreme UV Imager (EUI, Rochus et al. 2020) and the Spectral Imager of the Coronal Environment (SPICE, Anderson et al. 2020) spectrograph, can complement the coronal measurements of DKIST in various ways depending on the spacecraft-Earth configuration during the encounters. Quadrature windows can help disentangle line-of-sight effects from the DKIST Cryo-NIRSP coronal measurements (see Judge et al. 2013) as EUI will provide the missing context of the structures contributing to the line-of-sight average. Using the Cryo-NIRSP instrument in combination with SPICE will enable for the first time stereo-spectroscopy of outflows and plasma diagnostics, which are also heavily affected by line-of-sight effects. For these synergistic examples, field extrapolations from PHI can provide quantitative information of the magnetic fields contributing to the line-of-sight average.

Solar Orbiter conjunction configurations will provide an opportunity for simultaneous on-disk coordinated observations combining the Solar Orbiter remote sensing telescopes and the DKIST on-disk instrument suite, maximizing spectral coverage (from UV to the IR), temporal and spatial resolution, and sensitivity. Also, in conjunction, off-limb observations, including the coronagraph onboard Solar Orbiter (METIS; Antonucci et al. 2020), will benefit from coordinated observations with DKIST. Specifically, DKIST direct measurements of the magnetic field in the off-limb corona provide a unique opportunity to calibrate the indirect inferences of the closed vs. open magnetic configuration planned by METIS.

---

[3] *http://sci.esa.int/solar-orbiter/48985-solar-orbiter-definition-study-report-esa-sre-2011-14/*

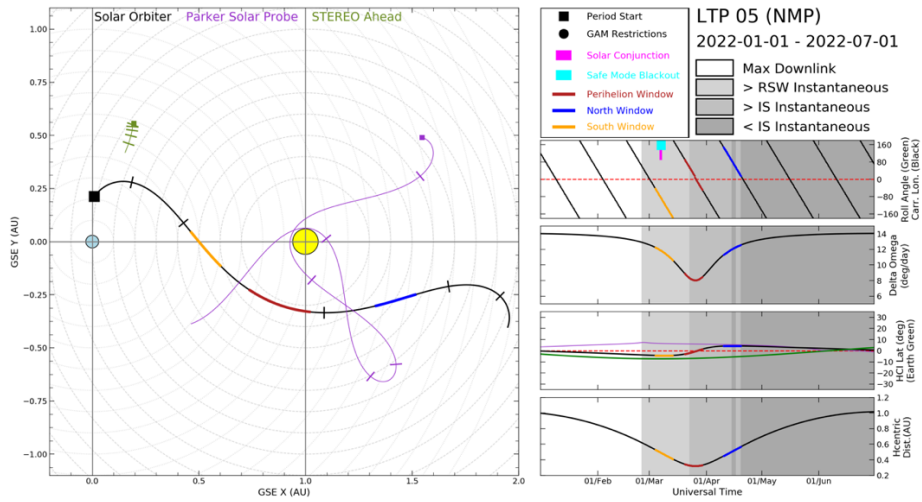

*Figure 2. Sample encounter orbit for the Solar Orbiter mission. The three remote sensing windows are indicated by colors (orange, dark red, blue). The orange window corresponds to a conjunction configuration and the dark red window to a quadrature encounter. Earth is at coordinates [0,0] on the left image, Sun at [1,0]. Image courtesy of ESA.*

The in-situ payload for Solar Orbiter includes a complete suite of instruments with analogs to those onboard PSP, but with some additional capabilities. The list consists of an energetic particle detector (EPD; Rodriguez Pacheco et al. 2020), a magnetometer (MAG; Horbury et al. 2020), a radio and plasma wave sensor (RPW; Maksimovic et al. 2020), and a solar wind analyzer (SWA; Owen et al. 2020). This last instrument is similar to the PSP's SWEAP suite, but with an added —NASA provided— key component, the Heavy Ion Sensor (HIS). SWA will give the solar wind composition measurements between 0.3 and 1 AU, including electrons, protons, alpha particles, but with HIS extending the composition analysis to heavier ions such as C, N, O, Si, Fe, S, and others. Measurements from SWA/HIS, in particular in combination with SPICE on-disk data, can help narrow down the connectivity options back to the Sun, as the charge-state ratios of heavy ions are an indication of the level of heating at the source and help distinguish among source candidates (coronal hole vs. streamer), although some modeling might be needed (e.g., Oran et al. 2015; Laming et al. 2019).

A key strategy to trace plasma parcels from the Sun into the heliosphere is to track the so-called First Ionization Potential (FIP)-bias, the anomalously high abundance (relative to the photosphere) of elements with low First Ionization Potential (see, e.g., Schmelz 2012). This effect is more pronounced in the slow solar wind than in the fast solar wind, which often shows photospheric abundances (Lee et al. 2015). The nature of the FIP effect is still under debate, but the first ionization of neutral elements occurs in the chromosphere, a layer that DKIST will map with unprecedented detail. Indeed, it is thought that new active regions emerge with photospheric abundances and develop the FIP-bias at chromospheric heights that then propagate into the corona (Laming 2015). This processing and propagation to the upper layers can give rise to a lag between magnetic activity indices and abundance fluctuations (Alterman and Kasper 2019). Comparing elemental abundances and FIP-bias inferred from remote sensing observations with in-situ counterparts is the main objective of the SPICE and SWA/HIS combination. Both instruments can determine the strength of the FIP-bias for several chemical elements. Coordinated observations tracking the same plasma packages are planned for several perihelion windows. This intrinsically multi-messenger approach to study the magnetic connectivity between the Sun and the heliosphere is one of the overarching goals of the Solar Orbiter mission. As is the case for PSP, by approaching the Sun to about 0.3 AU, Solar Orbiter will certainly facilitate tracing the plasma parcels back to the surface. DKIST can provide invaluable support by observing nearly simultaneously various layers of the Sun with different instruments and in distinct orbital configurations. It is now clear that the wavelength range covered by Cryo-NIRSP contains coronal lines which are potentially very unique for the FIP bias study (for instance from Ar XIII 834nm & 1014nm, Si IX 2.58μm & 3.93μm, S XIII 1030nm, S XI 1.39μm & 1.92μm, Fe XIII 1.07μm & 1.08μm; see Del Zanna & DeLuca, 2018). The inclusion of additional filters that observe such lines will complement the temperatures and elemental abundances coverage of the FIP bias studies from SPICE. Other visible lines from the ViSP instrument can provide additional diagnosis, but the coronal capabilities of this instrument need to be demonstrated first.

One area where DKIST can critically contribute to this investigation is by providing accurate magnetic field measurements during the period when Solar Orbiter is establishing the connectivity between the SPICE composition measurements and the SWA/HIS in-situ measurements. For orbital settings where the SPICE FOV is on the disk, DKIST can co-point and obtain vector magnetic maps in the photosphere and chromosphere. By adding the chromospheric magnetic information, DKIST can map the layer where composition (FIP bias) changes and where field line reconfiguration is expected to occur. A key target is a process known as interchange reconnection (Fisk & Schwadron, 2001; Crooker, 2002) whereby a closed field region interacts and reconnects with an open field line, establishing a new route for plasma to escape the Sun. The FIP bias occurs in the closed loop before interchange reconnection. After interchange reconnection, the FIP biased plasma is transmitted via the newly open channel into the solar wind (for a recent discussion of the factors at play, see Laming et al. 2019). It is crucial to understand if this reconnection is one of the main processes that give rise to the slow solar wind, and its identification and characterization require careful measurements of the changing connectivity. In quadrature encounters, DKIST will similarly seek to quantify the degree of field line openings occurring in coronal levels and the expansion rate of the resulting open field lines. We anticipate targeting both the near-limb region (DL-NIRSP) and the large-scale corona far from the limb (Cryo-NIRSP).

The coordinated observations described above have various potential target regions such as small-scale emergence episodes in the interior of coronal holes (e.g., Shimojo & Tsuneta, 2009) or the dynamics near hole boundaries as they have been consistently related with the slow solar wind (Ko et al., 2014). After the discovery of ubiquitous upflows near the edges of active regions by *Hinode* (Harra et al., 2008), these areas may constitute a source of the nascent slow solar wind. Using *Hinode*-estimated FIP bias ratios and in-situ composition data at 1 AU, Brooks & Warren (2011) traced back slow wind-like composition measurements to the edges of an AR that displayed outflows for at least five days. Thus, active-region boundaries represent another obvious target for coordinated observations between Solar Orbiter, *Hinode*, and DKIST. The additional out of the ecliptic vantage point provided by Solar Orbiter will produce new insights into the solar wind's source regions.

Coordination, including all three facilities, DKIST, PSP, and Solar Orbiter, also offers unique opportunities for synergistic science. The previous section already explained the case for combining remote sensing and in-situ Helium observations between DKIST and PSP. Bringing Solar Orbiter into the mix can prove keys to understanding the role of Helium in the various types of solar wind and the source regions. By using the EUI He II 304 Å channel in conjunction with Lyman-alpha observations from EUI and METIS, Solar Orbiter will be able to estimate the Helium abundance during the remote sensing windows. Coordinated observations with DKIST during on disk and off-limb passages using He I 1083 nm (and potentially other lines in the visible with the ViSP) will allow tracking the He (the highest FIP element) on the Sun from even deeper layers than what Solar Orbiter can do on its own. Combining DKIST and Solar Orbiter data during conjunction windows to map the presence of Helium in the solar corona while PSP measures in quadrature alpha particles from within 10 $R_\odot$ is a compelling way to understand the evolution of Helium in the nascent solar wind. It remains to be seen if the orbital mechanics of both missions will make possible such a joint campaign, in particular, towards the end of the PSP prime mission phase.

## 5. DKIST, PSP, Solar Orbiter, and spacecraft at 1 AU: Studying the evolution of CMEs observed in conjunction.

CME properties evolve as they propagate through the solar wind due to expansion and interaction with the solar wind (e.g., Bothmer & Schwenn 1996; Manchester et al. 2004; Lavraud et al. 2014; Ruffenach et al. 2015). A complete understanding of the changes in the properties of CMEs as they propagate in the inner heliosphere is required to understand their space weather impact at the Earth. Prior to PSP, the most recent in situ magnetic field and plasma observations of CMEs at < 0.7 AU have come from the MESSENGER spacecraft at Mercury (e.g., Winslow et al., 2015, 2016; Good et al., 2015). However, MESSENGER was a planetary mission, and although its observations were useful for heliophysics studies, they were quite limited. With the era of DKIST, PSP, and Solar Orbiter taking unprecedented measurements of the Sun and inner heliosphere environment, time is ripe to develop our physical understanding of the evolution and interaction of CMEs prior to reaching Earth, and thereby lay the groundwork for improved space weather forecasting. These three facilities will allow for unparalleled mapping of CMEs as they propagate from the Sun to 1 AU. For CMEs that are observed in conjunction, this can be achieved by DKIST providing accurate magnetic field measurements of the source as the CME erupts, PSP and/or Solar Orbiter taking in situ magnetic field and plasma measurements of the CME inside 1 AU, and STEREO/ACE/Wind/IMAP providing in situ measurements at 1 AU.

Even better monitoring will be achieved for CMEs that are observed in conjunction by both PSP and Solar Orbiter, providing two observations points in the innermost heliosphere, but in quadrature with respect to the Earth. This configuration allows the novel coronal DKIST observations to map the prominence-cavity system before it erupts. Coronal prominence-cavity systems are regions of stored magnetic energy that erupt in the form of CMEs. Already in quiescent conditions, the equilibrium state of the system is only partially understood. The inner prominence shows various types of small-scale dynamics which possibly induce material exchange with the cavity itself. High temperature and low density dominate the cavity. The models proposed for the cavity lack the constraints of magnetic field measurements, and only a few measurements of prominences magnetic field have been obtained (but see Wang et al. 2020). In this respect, DKIST will be able to provide accurate vector magnetic field measurements in both regions. The coordination with Solar Orbiter (using PHI and EUI) in quadrature will allow better constraint of the 3D magnetic structure and the changing topology from the chromosphere to the corona. This will also be possible thanks to the high latitudes reached by Solar Orbiter, which provides a completely new view of the cavity-prominence system. The dynamics (as well as other plasma diagnostics) of prominences and their interface with cavities will be obtained using the Doppler information from the SPICE spectra, together with DKIST in coronal mode. These studies will constrain eruption models with more quantitative and reliable information on the precursor state of the system.

## 6. Opportunities for Discovery

The previous sections showcased some of the combined science cases as opportunities for enhanced, synergistic science. Close coordination among the three facilities will lead to additional, often unanticipated opportunities. As a case example, DKIST, PSP, and Solar Orbiter will shed light into the so-called open flux problem (Linker et al. 2017), i.e., the persistent underestimation—by as much as a factor three—of the heliospheric magnetic flux predicted by models based on surface magnetograms, vs. what is measured in situ at 1 AU. First, it is unclear whether the open flux problem persists in the inner heliosphere. The FIELDS magnetometer onboard PSP will soon shed some light on this question. Second, the high resolution, high sensitivity vector magnetograms from DKIST will unveil the amount of flux that currently lies undetected in existing full-disk magnetograms, and that can be responsible for the larger flux observed at 1 AU.

Another case example is the lingering problem in solar and stellar astronomy of how stars create X-ray emitting coronae and winds. The processes leading to coronal heating are ultimately responsible for the escape of plasma from the Sun into the heliosphere. However, how the plasma is heated is still debated. A new generation of numerical models of the solar atmosphere, including MHD and radiative effects (Hansteen et al. 2015; see also Klimchuk 2006), have shown that the ever-present reconfiguration of chromospheric and coronal magnetic fields leads to the formation of magnetic discontinuities supporting the idea of nanoflares as the underlying process leading up to atmospheric heating. However, it is also known from high-resolution imaging and spectroscopy that the corona is permeated by MHD waves that can dissipate and heat the ambient coronal plasma (Cranmer & van Ballegooijen 2005; De Pontieu et al. 2007). It is almost certain that both processes, reconnection and wave dissipation, occur on the Sun in various forms and relevance depending on the specific region under study. Progressing in our understanding of the processes that dominate the heating and their underlying physics is a critical challenge in solar research where the multi-messenger approach represented by DKIST, PSP, and Solar Orbiter will provide a new perspective.

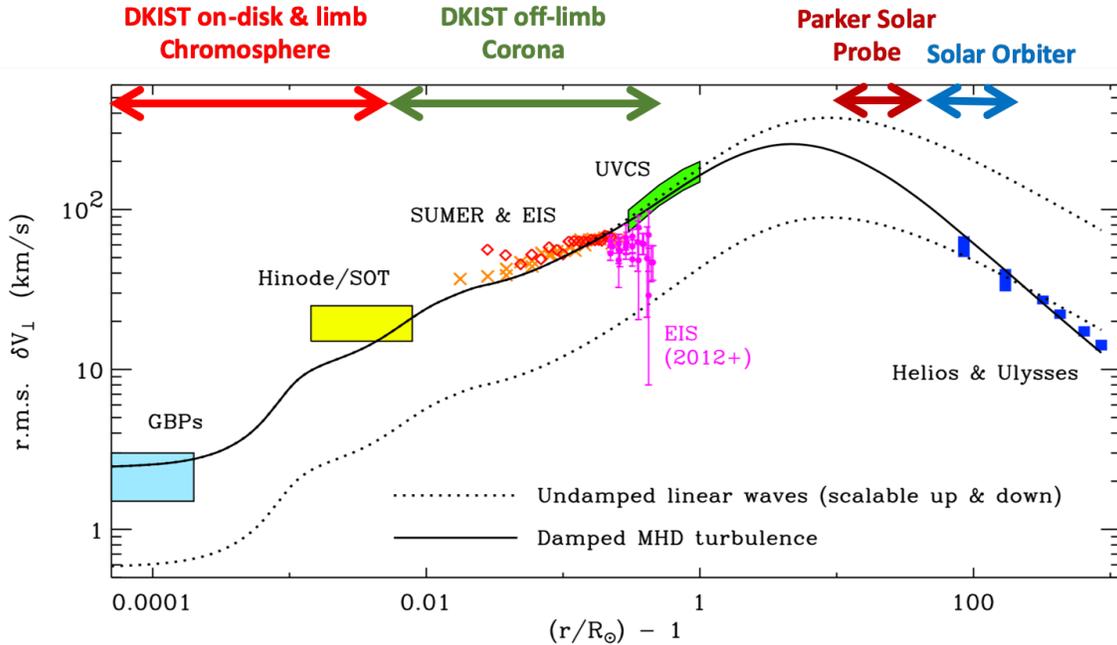

*Figure 3.* Velocity (rms) amplitude of transverse motions from the solar surface to the near-Earth environment. Squares and symbols inside one solar radius (above the limb) correspond to remote sensing observations as indicated. Blue squares are measurements from in-situ space missions. The solid and dashed lines correspond to damped and undamped Alfvénic wave turbulence (Cranmer & van Ballegooijen, 2005) models. The upper part of this figure indicates the regions accessible to DKIST observations (and, in some cases, Solar Orbiter remote sensing instruments) and to PSP and Solar Orbiter in-situ measurements. Adapted from Cranmer et al., 2017.

Figure 3 (see Cranmer et al. 2017) shows a clear example of how the multi-messenger constellation can synergistically produce constraints to the various physical processes that lead to heating processes. The lines in the figure indicate the magnitude of the transverse motions present in a model of damped (solid) and undamped (dashed) Alfvén wave turbulence compatible with solar remote sensing and in-situ observations (Cranmer & van Ballegooijen 2005; Crammer et al. 2017). The figure starts with the photospheric transverse motions observed in the so-called G-band bright points (GBPs). These transverse motions are thought to be responsible for shaking magnetic field lines leading to upward propagating Alfvén waves. While these GBPs motions have been measured using intensity proxies for decades, very few (magnetic) wave power estimates are available (but see Jafarzadeh et al. 2013). DKIST will perform much more accurate and quantitative measurements of these photospheric motions and their magnetic consequences, as demonstrated recently by Van Kooten & Cranmer (2017). The *Hinode*/Solar Optical Telescope (SOT) box in yellow indicates the amplitude of the transverse motions detected by *Hinode* using Ca II H chromospheric images of type II spicules (De Pontieu et al. 2007). Using DKIST's higher spatial resolution and magnetic sensitivity, we will be able to quantify the energy fluxes carried out by these waves by measuring the perturbations directly in the magnetic field lines. Critical instruments for estimating the induced velocity and the magnetic fluctuations are the DL-NIRSP and the Visible Tunable Filter (VTF). Moving away from the solar surface in Figure 3, the next set of measurements corresponds to nonthermal line widths observed by instruments in Solar and Heliospheric Observatory (SOHO) and *Hinode*. They reach out to 1 R$_\odot$ above the surface and display mixed indications of damping processes. SOHO's UltraViolet Coronagraph Spectrometer (UVCS) (green square) shows little evidence of damping. However, more recent measurements by *Hinode's* Extreme-ultraviolet Imaging Spectrometer (EIS) (magenta dots in the figure) show signs of damping mechanisms acting in the corona that contrast with the UVCS analyses. It is urgent to confirm the existence, or not, of these damping processes and consolidate our understanding of how Alfvén waves can deposit their energy and convert it to heat in this inner region of the corona. Cryo-NIRSP routine observations of spectral lines in this range of heights will undoubtedly provide the observational constraints needed to clarify the amount of damping present in this region of the solar corona. Solar Orbiter coordinated observations in conjunction windows can provide nonthermal line widths of other species and create a complete picture of the damping processes at play in the solar corona. The early phases of DKIST operations should aim at including coordinated campaigns with *Hinode*/EIS to ensure cross-calibrations of past and future non-thermal doppler signals.

The models in Figure 3 reach out past the Earth and is contrasted with the observed in-situ magnetic field fluctuating amplitudes detected by Helios and Ulysses. A good match with existing data (blue squares) requires the presence of wave damping in the heliosphere (solid line). The instruments onboard PSP and Solar Orbiter are going to provide much better coverage over a broader range of distances to the Sun and in a variety of solar and heliospheric structures. Fine tuning of models like those used in Figure 3 to fit the various observations from DKIST/Solar Orbiter and the in-situ detections from PSP/Solar Orbiter will fundamentally constrain the potential physical processes behind wave heating models.

## 7. Concluding Remarks

This work is intended to describe the unique scientific opportunities enabled by NSF's DKIST capabilities —especially for observing the corona— in coordination with the PSP and Solar Orbiter missions leading to an unprecedented multi-messenger era for Solar and Heliospheric Physics. As demonstrated, this work points towards the need for coordination between NSF, NASA, and the ESA partners to maximize the scientific return that this unique constellation offers. Special calls for coordinated analysis that the community could use to facilitate the science goals described in this document will help alleviate the historical divide between solar space and ground base communities and funding streams.